\documentclass[a4paper,11pt]{article}
\usepackage{latexsym}
\usepackage{epsfig, subfigure, subfloat, graphicx, float }
\usepackage{amssymb}
\usepackage{amsmath}
\author{H. Mohseni Sadjadi \footnote{mohsenisad@ut.ac.ir} and Parviz Goodarzi
\\ {\small Department of Physics, University of Tehran,}
\\ {\small P. O. B. 14395-547, Tehran 14399-55961, Iran}}
\title{Reheating temperature in non-minimal derivative coupling model}
\begin{document}
\maketitle
\begin{abstract}
We consider the inflaton as a scalar field described by a
non-minimal derivative coupling model with a power law potential. We
study the slow roll inflation, the rapid oscillation phase, the
radiation dominated and the recombination eras respectively, and
estimate e-folds numbers during these epochs. Using these results
and recent astrophysical data we determine the reheating temperature
in terms of the spectral index and the amplitude of the power
spectrum of scalar perturbations.
\end{abstract}

\section{Introduction}

To solve some dilemmas in the standard model of cosmology such as the
flatness, the horizon, the monopoles problems and so on,
inflation as an accelerated expansion era in the early universe
was introduced by \cite{inf}. This scenario is now dubbed as old model of
inflation, in which the universe underwent a de-Sitter expansion
in a supercooled unstable false vacuum. Afterwards,  by proposing
a scalar field (inflaton) as the source of inflation, a new
inflationary model was introduced in \cite{lind}. In this context,
inflation was driven by the inflaton which slowly rolled down
towards the minimum of its effective potential. To provide enough
e-folds number, the potential must be nearly flat near its
minimum.

The nature of this scalar field has not yet been identified, but a
simple possible candidate might be the Higgs boson \cite{Higgs}. To
adapt the inflaton to the Higgs boson, a non-minimally derivative
coupling model in which the kinetic term of the inflaton is coupled
to the Einstein tensor, was proposed in \cite{germ1}. This model
does not suffer from unitary violation problem and is safe from
quantum corrections. Besides, slow roll inflation can be described
by steep potentials in this framework\cite{germ2,germ3}. More
general non-minimal derivative coupling model has also been
considered in the literature to study the accelerated expansion of
the universe in the early universe as well as in the late time
\cite{non}.

After the end of the inflation, the universe was cold and
dominated by the inflaton scalar field energy. This energy had to
be converted to relativistic particles to reheat the universe via
a procedure called the reheating process \cite{reh}. A proposal for
reheating, is the decay of the inflaton to ultra-relativistic
particles (radiation) during a rapid coherent oscillation phase
about the minimum of the potential. These particles interacted
rapidly to become in thermal equilibrium characterized by the
reheating temperature $T_{reh}$, and the universe entered on the
radiation dominated era.

Although the exact value of $T_{reh}$ has not yet been known, but some
upper and lower bounds for this temperature have been obtained in
the literature. By considering that the reheating process occurred
before the big bang nucleosynthesis (BBN), and by combining constraints
on light elements abundance and data obtained from large scale structure and
cosmic microwave background (CMB), one can find a lower bound  for
$T_{reh}$, $4 MeV \lesssim T_{reh}$ \cite{BBN}. An upper bound may
be taken as the energy scale at the end of inflation which is
around the GUT scale $T_{reh}\lesssim 10^{16}GeV$. These
assumptions give a wide range for the reheating temperature. In \cite{Martin}
by involving supersymmetry and considering the gravitino
production, and on the base of cosmic microwave background (CMB)
radiation data, this range was tightened to $6TeV \lesssim
T_{reh}\lesssim 10^{4}TeV$.

A more accurate method to determine $T_{reh}$ in terms of CMB data
was introduced in \cite{Mielc}. This method is based on determining the
number of e-folds during the evolution of the universe from the
inflation until the present time. Although in this context
$T_{reh}$ may be determined in terms of spectral index and
amplitude of power spectrum of scalar perturbations, but due to
uncertainties of theses quantities in WMAP7 data \cite{WMAP}, a large relative uncertainty
for $T_{reh}$ is arisen: ${\sigma(T_{reh})\over T_{reh}}\approx
53$, where $T_{reh}=3.5\times10^6GeV$.

In this paper we assume that the inflaton is a scalar field
described by non-minimal derivative coupling model introduced in
\cite{germ1}. We study the inflationary era, the rapid oscillation
phase of the inflaton, the radiation dominated epoch respectively
and employ the method proposed in \cite{Mielc} to determine the
reheating temperature in terms of the spectral index and power
spectrum of scalar perturbations. Finally, the value of $T_{reh}$
and its relative uncertainty are computed from recent
astrophysical data such as WMAP9 and Planck 2013 results.

We use units $\hbar=c=1$ through the paper.

\section{Evolution of the universe and the reheating temperature }
We consider the spatially flat Friedmann-Lema\^{\i}tre-
Robertson-Walker (FLRW) space-time
\begin{equation}\label{1}
ds^2=-dt^2+a^2(t)(dx^2+dy^2+dz^2),
\end{equation}
and choose an arbitrary length scale, $\lambda_0$, crossing the
Hubble radius $R_H:={1\over H}={a\over \dot{a}}$ at some time,
denoted by $t_{*}$, during the inflation \cite{large}. By using the red-shift
relation
\begin{equation}\label{2}
{\lambda(t_{*})\over \lambda_0}={a(t_{*})\over a_0},
\end{equation}
where subscript $"0" $ denotes the present time and by taking
$a_0=1$, we obtain
\begin{equation}\label{3}
\lambda_0={1\over a(t_{*})H(t_{*})}.
\end{equation}
This reference time will be used in division of the evolution of
the universe into four parts as follows:

I- From $t_{*}$  until the end of slow roll, denoted by $t_e$.

II- From $t_e$ until the reheating or beginning of the radiation
dominated epoch, denoted by $t_{reh}$.

III- From $t_{reh}$ until recombination era, denoted by $t_{rec}$.

IV- From $t_{rec}$ until the present time $t_0$.

The number of e-folds from $t_{*}$ until $t_0$ is then given by
\begin{eqnarray}\label{4}
\mathcal{N}=\ln\left({a_0\over a_{*}}\right)&=&\ln\left(({a_0\over
a_{rec}})({a_{rec}\over a_{reh}})({a_{reh}\over
a_{e}})({a_{e}\over a_{*}})\right)\nonumber \\
&=&\mathcal{N}_{IV}+\mathcal{N}_{III}+\mathcal{N}_{II}+\mathcal{N}_{I},
\end{eqnarray}
where the subscripts denote the value of the parameter at their
corresponding times.

In the following we will try to use eq.(\ref{4}) to determine the
reheating temperature in a non-minimal derivative coupling model
in which the inflaton kinetic term is non-minimally coupled to
Einstein tensor. This inflationary model is described by the
action \cite{germ1}
\begin{equation}\label{5}
S_{\phi}=\int d^4x\sqrt{-g}\left[{M_P^2\over 2}R-{1\over 2}g^{\mu
\nu}\partial_\mu \phi\partial_\nu \phi+{1\over 2M^2}G^{\mu
\nu}\partial_\mu \phi \partial_\nu \phi -V(\phi)\right],
\end{equation}
where $G^{\mu \nu}=R^{\mu \nu}-{1\over 2}g^{\mu \nu}R$ is the
Einstein tensor, $R$ is the scalar curvature, $M_P$ is the
reduced planck mass given by $M_P=\sqrt{{1\over 8\pi G}}=2.4\times
10^{18} GeV$, and $M$ is a scale with mass dimension.

\subsection{Slow roll}

In the era (I), the universe is dominated by the inflaton scalar
field. The Friedmann equation is
\begin{equation}\label{6}
H^2={1\over 3M_P^2}\rho_\phi,
\end{equation}
where
\begin{equation}\label{7}
\rho_{\phi}={1\over 2}\left(1+9{H^2\over
M^2}\right)\dot{\phi}^2+V(\phi),
\end{equation}
is the energy and the upper dot is $"^." ={d\over dt}$. The pressure is obtained as
\begin{equation}\label{8}
P_{\phi}={1\over 2}\left(1-3{H^2\over
M^2}\right)\dot{\phi}^2-V(\phi)-{1\over M^2}{d(H\dot{\phi}^2)\over
dt}.
\end{equation}
With the help of continuity equation
\begin{equation}\label{9}
\dot{\rho_{\phi}}+3H(P_{\phi}+\rho_{\phi})=0,
\end{equation}
one can derive the equation of motion of the inflaton as
\begin{equation}\label{10}
\left(1+{3H^2\over M^2}\right)\ddot{\phi}+3H\left(1+{3H^2\over
M^2}+{{2\dot{H}}\over M^2}\right)\dot{\phi}+V'(\phi)=0.
\end{equation}

In the continue, we restrict ourselves to high friction regime \cite{germ1}
\begin{equation}\label{11}
{H^2\over M^2}\gg1.
\end{equation}
This choice, as we will see, by enhancing the slow roll procedure,
enables us to consider more general steep potentials. During the
slow roll, we have  ${H^2\over M^2}\dot{\phi}^2\ll V(\phi)$,
$\ddot{\phi}\ll H\dot{\phi}$, and therefore the Friedmann equation
and inflaton equation of motion reduce to
\begin{eqnarray}\label{12}
&&H^2\approx {1\over 3M_P^2}V(\phi),\nonumber \\
&&\dot{\phi}\approx -{M^2V'(\phi)\over 9H^3}
\end{eqnarray}
respectively.  The slow roll parameters $\epsilon$, $\delta$
satisfy
\begin{equation}\label{13}
\epsilon:=-{\dot{H}\over H^2}\simeq {M_P^2\over 2}{M^2\over
3H^2}{V'^2(\phi)\over V^2(\phi)}\ll1,\,\,\delta:={\ddot{\phi}\over
H\dot{\phi}}\ll 1.
\end{equation}
$\delta$ can be written as
\begin{equation}\label{14}
\delta\simeq -\eta+3\epsilon
\end{equation}
where $\eta$ is defined with
\begin{equation}\label{15}
\eta:={M_P^2\over 3}{M^2\over
H^2}{V''(\phi)\over V(\phi)}.
\end{equation}
Eqs.(\ref{13}-\ref{15}), imply that by choosing an appropriate $M$ in high friction
regime (\ref{11}), slow roll conditions do not oblige us to adopt
approximately flat potentials.

Hereafter we will restrict ourselves to power law potential
\begin{equation}\label{16}
V(\phi)=v \phi^n,
\end{equation}
where $v$ is a real number, and $n$ is an even positive integer to
guarantee that the potential has a minimum, about which the rapid
oscillation of the inflaton occurs after slow roll.

The number of e-folds in the era (I) is
\begin{eqnarray}\label{17}
\mathcal{N}_{I}=\ln\left({a_e\over a_{*}}\right)=\int_{t_*}^{t_e}Hdt&=&{1\over M_P^4M^2}\int_{\phi_e}^{\phi_{*}}{V^2(\phi)\over V'(\phi)}d\phi\nonumber \\
&\simeq& {v\over n(n+2)M^2M_P^4}\phi_{*}^{n+2}.
\end{eqnarray}
To obtain the above equation we have used (\ref{16}), and $\phi_*\gg \phi_e$. To estimate
$\phi_{*}$, we consider the spectral index $n_s$ \cite{germ2},
\begin{equation}\label{18}
n_s-1\simeq -2\epsilon -2\delta\approx {M_P^2M^2\over
H_{*}^2}\left[-{4\over 3}{V'^2(\phi_{*})\over
V^2(\phi_{*})}+{2\over 3}{V''(\phi_{*})\over V(\phi_{*})}\right].
\end{equation}
For the power law potential, this equation reduces to
\begin{equation}\label{19}
1-n_s={2M_P^4M^2n(n+1)\over v}\phi_{*}^{-(n+2)}.
\end{equation}
The number od e-folds in the time interval I is then obtained by
substituting $\phi_{*}$ from (\ref{19}) into (\ref{17}):
\begin{equation}\label{20}
\mathcal{N}_{I}=2{(n+1)\over (n+2)(1-n_s)}.
\end{equation}
\subsection{Reheating era}
\begin{figure}[h]
\centering\epsfig{file=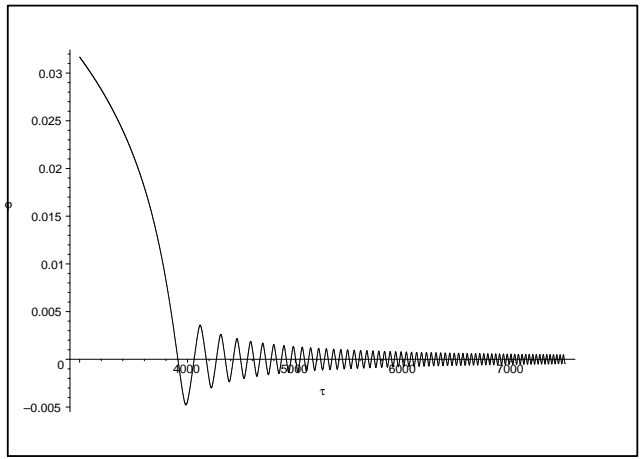,width=6cm,angle=270}
\caption{${\varphi:={\phi\over M_P}}$ in terms of dimensionless
time $\tau=mt$, for ${m^2\over M^2}=10^{8}$ with initial conditions
$\{\varphi(1)=0.056$, $\dot{\varphi}(1)=0\}$, for the quadratic
potential. } \label{fig1}
\end{figure}

At the end of slow roll we have $\epsilon(\phi_e)\simeq 1$, which
yields
\begin{equation}\label{21}
\phi_e^{n+2}={M_P^4M^2n^2\over 2v}.
\end{equation}
At this time the energy density is approximately given by
\begin{equation}\label{22}
\rho_e\simeq V(\phi_e)=v\left({M^2M_P^4n^2\over 2v}\right)^{n\over
n+2},
\end{equation}
and the scalar field commences a rapid coherent oscillation around
the bottom of the potential (see fig.(\ref{fig1}), depicted for the
quadratic potential $V(\phi)={1\over 2}m^2\phi^2$ via numerical
methods).
To obtain the equation of state of the universe in this era, we
follow the steps used in \cite{rapid}.  In this high frequency
regime, the behavior of the scalar field is opposite to the slow
roll, and its quasi-periodic evolution may be described as
\cite{rapid,god}
\begin{equation}\label{r1}
\phi(t)=\Phi(t)\cos\left(\int W(t)dt\right).
\end{equation}
$W(t)$ is some function of time and the time dependent amplitude ,
$\Phi(t)$, is given by
\begin{equation}\label{r2}
V(\Phi(t))=v\Phi^n(t)=\rho_\phi.
\end{equation}
The rapid oscillation of the scalar field $\phi$ occurs after the
slow roll. This high frequency regime is characterized by (for more
details, see the first reference in \cite{rapid} and also
\cite{god})
\begin{equation}\label{r3}
\left|{\dot{\Phi(t)}\over \Phi(t)}\right|=\left|{2\over
n}{\dot{H}\over H}\right|=\left|{1\over n}{\dot{\rho_\phi}\over
\rho_\phi}\right|\ll W(t),
\end{equation}
implying that the energy density and the Hubble parameter decrease
very slowly (insignificantly) in one period of oscillation of the
scalar field \cite{rapid}. This can be seen numerically in fig
.(\ref{fig1}) which shows that although $\phi$ oscillates rapidly,
but ${\delta\Phi(t)\over \Phi(t)}\ll 1$, where $\delta\Phi(t)$ is
the change of $\Phi(t)$ during one period of $\phi$ oscillation.

\begin{figure}[h]
\centering\epsfig{file=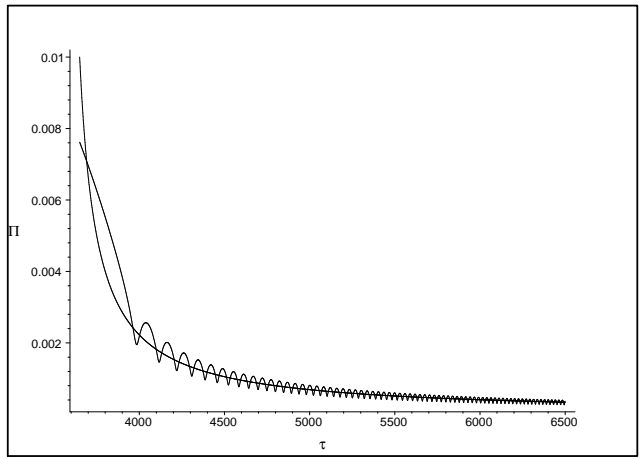,width=6cm,angle=270}
\caption{${\Pi:={H\over m}}$ in terms of dimensionless time $\tau=m
t$,  with initial conditions $\{\varphi(1)=0.056$,
$\dot{\varphi}(1)=0\}$, where $\varphi={\phi\over M_P}$, for the
quadratic potential ${1\over 2}m^2\phi^2$ and ${m^2\over
M^2}=10^{8}$. } \label{fig2}
\end{figure}

With the help of (\ref{7}), (\ref{8}), and (\ref{11}), the
continuity equation of the scalar field can be written as
\begin{equation}\label{r4}
\dot{\rho_{\phi}}=-2H\left(\rho_{\phi}-V(\phi)\right)+{3H\over
M^2}{d\over dt}\left(H\dot{\phi}^2\right).
\end{equation}
Now let us take the time average of both sides of the above equation
over one oscillation of the scalar field( $<...>=
{\int_t^{t+T}...dt'\over T}$ is the time average over an oscillation
whose period is $T$). The left hand side of (\ref{r4}) gives
\begin{equation}\label{r5}
<\dot{\rho_{\phi}}>={\int_t^{t+T}\dot{\rho_{\phi}}dt'\over
T}={\delta \rho_{\phi(t)}\over T}\approx \dot{\rho_{\phi}}(t).
\end{equation}
This approximation is valid only on time scales large with respect
to the short period of high frequency quasi-oscillation. Converting
time integration to $\phi$ integration and using
\begin{equation}\label{r6}
\dot{\phi}^2={2M^2\over 9H^2}(\rho_{\phi}-V(\phi)),
\end{equation}
and (\ref{r2}), we obtain the time average over an oscillation of
the right hand side of (\ref{r4}) as
\begin{eqnarray}\label{r7}
&&{-2H\int_{-\Phi}^{\Phi}\sqrt{\rho_\phi-V(\phi)}d\phi\over
\int_{-\Phi}^{\Phi} {d\phi\over \sqrt{\rho_\phi-V(\phi)}}}+{2\over
3T}(\rho_\phi-V(\phi))\Big|^\Phi_{-\Phi}\nonumber \\
&\approx&-2v H\int_{-\Phi}^{\Phi}\sqrt{\Phi^n-\phi^n}d\phi\over
\int_{-\Phi}^{\Phi} {d\phi\over \sqrt{\Phi^n-\phi^n}}\nonumber \\
&\approx&-2v {n\over n+2}H\Phi^n.
\end{eqnarray}
To obtain the above relation we have used the evenness of $V$,
(\ref{r2}), and also (\ref{r3}) which implies that the variables in
the integral except $\phi$ may be replaced by their averaged values
in one oscillation of $\phi$ in rapid oscillation phase. So finally
we get
\begin{equation}\label{r8}
\dot{\rho_\phi}\approx {-2n\over n+2}H\rho_\phi.
\end{equation}
In (\ref{r7}) and (\ref{r8}), $\rho_\phi(t)$ and $H$ are the
averaged values in the sense explained above and (\ref{r8}) holds
for time scales large with respect to $T$ ($t\gg T$). Therefore on
time scales much larger than the period of rapid oscillation, the
effective equation of state parameter of the scalar field is
$w=\gamma-1$ where
\begin{equation}\label{r9}
\gamma={2n\over 3n+6}.
\end{equation}
If the non-minimal derivative coupling was absent we would have
$\gamma={2n\over n+2}$ \cite{rapid}. In this regime, following our
analysis and by using (\ref{r8}), the Hubble parameter can be
approximated by \cite{god}
\begin{equation}\label{25}
H=\sqrt{{1\over 3M_P^2}\rho_{\phi}}\simeq {2\over 3\gamma t}.
\end{equation}
To confirm and justify our results,  based on eqs.(\ref{6}),
(\ref{7}), and (\ref{10}), the behavior of $H(\propto
\sqrt{\rho_\phi})$ is depicted via numerical methods in
fig.(\ref{fig2}). This shows a good agreement between numerical
result and (\ref{25}) for large time scales.

We assume that in this epoch the inflaton decays to
ultra-relativistic particles (whose the energy density is denoted by
$\rho_r$), to reheat the universe. From the beginning of rapid
oscillation, i.e. $\rho_r=0$, until $\rho_r=\rho_{reh}\simeq
\rho_\phi$, which is the beginning of radiation dominated era, the
universe is approximately dominated by the rapidly oscillating
scalar field. Therefore, in this era the Hubble parameter can be
approximated by (\ref{25}) \cite{god}, and the scale factor during
this era scales as
\begin{equation}\label{26}
{a_{reh}\over a_{e}}\simeq\left({\rho_e\over
\rho_{reh}}\right)^{n+2\over 2n}.
\end{equation}
At $t_{reh}$ we can estimate the energy density as
\begin{equation}\label{27}
\rho_{reh}\simeq {g_{reh}\over 30}\pi^2T_{reh}^4,
\end{equation}
 where $T_{reh}$ is the temperature of ultra relativistic
 particles at $t_{reh}$, and $g_{reh}$ is the number of (massless) degrees of freedom corresponding to the
 ultra-relativistic particles present in the model at $t_{reh}$ \cite{kolb}.
 Collecting all together, we can estimate the number of e-folds
 during rapid oscillation
\begin{eqnarray}\label{28}
\mathcal {N}_{II}&=&\ln\left({{a_{reh}\over
a_{e}}}\right)\nonumber \\
&=&\ln\left(\sqrt{M_P^4M^2n^2\over 2}\left({30\over
g_{reh}\pi^2T_{reh}^4}\right)^{n+2\over 2n}v^{1\over n}\right).
\end{eqnarray}

To be more specific we must determine $v$. To do so, consider the
power spectrum of the scalar perturbation
\begin{equation}\label{29}
\mathcal{P}_s={H^2\over 8\pi^2M_P^2\epsilon},
\end{equation}
which is computed at the horizon crossing $k=k_0=a_{*}H_{*}$ (see (\ref{2}) and (\ref{3})),
where $k_0={1\over \lambda_0}$ is a pivot scale.  Thus
\begin{equation}\label{30}
H_*=2\pi M_P\sqrt{2\epsilon\mathcal{P}_s(k_0)}.
\end{equation}
Using this equation together with (\ref{19}) and $H_*^2\simeq {1\over 3M_P^2}V(\phi_*)$, and after some computations we find that
\begin{equation}\label{31}
v=\left(1-n_s\over 1+n\right)^{1+n}\left(6\pi^2
\mathcal{P}_s(k_0)\right)^{n+2\over 2}\left(2M^2\right)^{-{n\over
2}}M_P^4n.
\end{equation}
Substituting (\ref{31}) into (\ref{28}) yields
\begin{equation}\label{32}
\mathcal {N}_{II}=\ln\left[{1\over 2}\left({n(1-n_s)\over
1+n}\right)^{1+n\over n}\left({180 M_P^4\mathcal{P}_s(k_0)\over
g_{reh}T_{reh}^4}\right)^{n+2\over 2n}\right].
\end{equation}

So far, in our computations we have implicitly assumed that
${H^2\over M^2}\gg 1$ holds during inflation and reheating. But as
the inflaton oscillation amplitude and consequently $H^2$ (see
eq.(\ref{r2})) decrease, the validity of this assumption must be
investigated. At the end of the reheating era (beginning of
radiation dominated era) $\rho_\phi\simeq \rho_{reh}$, we have
$H_{reh}^2\simeq {1\over 3M_P^2}\rho_{reh}$. As $H$ decreases,
$H_{reh}$ is the the minimum of the Hubble parameter in the era I
and II; so if
\begin{equation}\label{r10}
{H_{reh}^2\over M^2}\gg 1,
\end{equation}
then the assumption ${H^2\over M^2}\gg 1$ is safe in our
computations. In terms of the reheating temperature, (\ref{r10})
may be written as
\begin{equation}\label{r11}
{\pi^2 g_{reh}\over 90 M^2 M_P^2}T_{reh}^4\gg 1.
\end{equation}

\subsection{Radiation dominated and recombination eras} In the
radiation dominated era the universe is constituted of
ultra-relativistic particles which are in thermal equilibrium, and
undergoes an adiabatic expansion during which the entropy per
comoving volume is conserved: $dS=0$ \cite{kolb}. In this era the
entropy density, defined by $s=Sa^{-3}$, is obtained as \cite{kolb}
\begin{equation}\label{33}
s={2\pi^2\over 45}g_{reh}T^3.
\end{equation}
So we have
\begin{equation}\label{34}
{a_{rec}\over a_{reh}}={T_{reh}\over T_{rec}}\left({g_{reh}\over
g_{rec}}\right)^{1\over 3}.
\end{equation}
In the recombination era, $g_{rec}$ corresponds to photons degrees of freedom and consequently $g_{rec}=2$. Hence
\begin{equation}\label{35}
\mathcal{N}_{III}=  \ln\left({T_{reh}\over
T_{rec}}\left({g_{reh}\over 2}\right)^{1\over 3}\right).
\end{equation}

The temperature decreases by the expansion of the universe via
$T(z)=T(z=0)(1+z)$, where $z$ is the redshift parameter. Therefore we can express $T_{rec}$ in terms of $T_{CMB}$ as
\begin{equation}\label{36}
T_{rec}=(1+z_{rec})T_{CMB}.
\end{equation}
We have also
\begin{equation}\label{37}
{a_0\over a_{rec}}=(1+z_{rec})
\end{equation}
Thus
\begin{equation}\label{38}
\mathcal{N}_{III}+\mathcal{N}_{IV}=\ln\left({T_{reh}\over
T_{CMB}}\left({g_{reh}\over 2}\right)^{1\over3}\right).
\end{equation}
\subsection{Reheating temperature}
So far, we have determined the number of e-folds in the the right
hand side of (\ref{4}). To obtain the reheating temperature we need also
to determine  $\mathcal{N}$ in (\ref{4}). Taking $a_0=1$, the number of
e-folds from the horizon crossing until the present time is
derived as $\mathcal{N}=\exp(\Delta)$, where
\begin{equation}\label{39}
\Delta={1\over a_*}={H_*\over k_0}={2\pi M_P\over
k_0}\sqrt{{2n(1-n_s)\over {1+n}}\mathcal{P}_s(k_0)}.
\end{equation}
To obtain the above relation we have made use of (\ref{30}) and
\begin{equation}\label{40}
\epsilon\simeq {n(1-n_s)\over 4(n+1)},
\end{equation}
which is derived from (\ref{13}-\ref{15}) and (\ref{19}).

Using (\ref{4}), (\ref{20}), (\ref{32}), (\ref{38}), and (\ref{39}) we finally arrive at

\begin{equation}\label{41}
T_{reh}=\alpha T_{CMB}^{-{n\over n+4}},
\end{equation}
where $\alpha$ is defined through
\begin{eqnarray}\label{42}
\alpha&=&g_{reh}^{n+6\over 6n+24}M_P\left({k_0 \over 2^
{11\over 6}\pi}\right)^{{n\over n+4}}\left({ 180n(1-n_s)\over
n+1}\right)^{n+2\over 2n+8}\times\nonumber \\
&&\exp\left({2n(n+1)\over (n+2)(n+4)(1-n_s)}\right)
\mathcal{P}_s^{1\over n+4}(k_0).
\end{eqnarray}
$\alpha$, up to this order of approximation, is independent of $v$
and $M$.  Note that, following (\ref{r11}), validity of the high
friction assumption (\ref{11}) requires that $M$ satisfies
\begin{equation}
M^2\ll {\pi^2 g_{reh}\over 90 M_P^2}\alpha^2T_{CMB}^{-4n\over n+4}.
\end{equation}

For the quartic potential, $n=4$, (\ref{41}) reduces to
\begin{equation}\label{44}
T_{reh}=1.927g_{reh}^{-{5\over 24}}M_P(1-n_s)^{3\over
8}\mathcal{P}_s^{1\over 8}(k_0)\exp\left({5\over 6(1-n_s)}\right)
\left({k_0\over T_{CMB}}\right)^{1\over 2},
\end{equation}
and for quadratic potential, $n=2$, it reduces to
\begin{equation}\label{46}
T_{reh}=2.205g_{reh}^{-{2\over
9}}M_P(1-n_s)^{1\over 3}\mathcal{P}_s^{1\over
6}(k_0)\exp\left({1\over 2(1-n_s)}\right) \left({k_0\over
T_{CMB}}\right)^{1\over 3},
\end{equation}
In the minimal model, for the quadratic potential, the reheating
temperature was obtained in \cite{Mielc}:
\begin{equation} \label{48}
T_{reh}=0.017M_P(1-n_s)^{1\over 2}\mathcal{P}_s^{-{1\over
2}}(k_0)\exp\left({6\over 1-n_s}\right)\left({k_0\over
T_{CMB}}\right)^{3}.
\end{equation}
Note that in contrast to the non-minimal derivative coupling model,
$T_{reh}$ in (\ref{48}) does not depend on relativistic degrees of
freedom $g_{reh}$.

By taking $g_{reh}=106.75$, which is the ultrarelativistic degrees
of freedom at the electroweak energy scale, the reheating
temperature in minimal model for quadratic potential was computed in
\cite{Mielc} as $T_{reh}=3.5\times10^{6}GeV$. This result was based
on WMAP7 data \cite{WMAP} which, for the pivot scale
$k_0=0.002Mpc^{-1}$, imply (for $\%68$ CL, or $1\sigma$ error)
\begin{eqnarray}\label{43}
\mathcal{P}_s(k_0)&=&2.441^{+0.088}_{-0.092}\times10^{-9}\nonumber \\
n_s&=&0.963\pm 0.012.
\end{eqnarray}
The relative uncertainty, ${\sigma(T_{reh})\over T_{reh}}$ up to
first order Taylor expansion, where
\begin{equation}\label{2000}
\sigma(T_{reh})=\sqrt{\left({\partial T_{reh}\over
n_s}\right)^2\sigma^2(n_s)+\left({\partial T_{reh}\over
\mathcal{P}_s}\right)^2\sigma^2(\mathcal{P}_s)}
\end{equation}
was derived as ${\sigma(T_{reh})\over T_{reh}}\approx 53$
\cite{Mielc}.  In our nonminimal model, for quadratic potential, and
by using (\ref{43}), we find $T_{reh}=6.53\times10^{12}GeV$ which is
much larger than what was obtained in the minimal case. The relative
uncertainty is now
\begin{equation}
{\sigma(T_{reh})\over T_{reh}}=4.275.
\end{equation}

Due to exponential dependence of $T_{reh}$ on $n_s$, the uncertainty
in determining the spectral index has a large effect on the
reheating temperature uncertainty. Fortunately recent results from
WMAP9, ACT, SPT and Planck 2013, may be employed to obtain more
exact value for reheating temperature with less relative
uncertainty. These results are quoted in the table1, {\it{for $\%68$
CL, or $1\sigma$ error}}. The adopted pivot scale for the two first
column is $k_0=0.002Mpc^{-1}$, while for the two last columns
$k_0=0.05Mpc^{-1}$.

\begin{table}[h]
\caption{Reheating temperature and its relative uncertainty}
\begin{tabular}{|c cccc|}
  \hline
   &$WMAP9$& $WMAP9+eCMB$& $Planck(only)$&$Planck+WP$ \\
   &       & $+BAO+H_0$   &               &$+highL+BAO$\\
  \hline
  $n_s$ & $0.972\pm0.013$ & $0.9608\pm0.0080$ & $0.9616\pm0.0094$ & $0.9608\pm0.0054$ \\
  $10^9\mathcal{P}_s$ & $2.41\pm0.1$ & $2.464\pm0.072$ & $2.23\pm0.16$ & $2.200\pm0.056$ \\
  $T_{reh(n=2)}(GeV)$ & $4.57\times10^{14}$ & $3.12\times10^{12}$ & $1.16\times{10}^{13}$ & $8.96\times10^{12}$ \\
   $T_{reh(n=4)}(GeV)$ & $2.42\times10^{15}$ & $5.58\times10^{11}$ & $4.26\times10^{12}$ & $2.75\times10^{12}$ \\
  $[{\sigma(T_{reh})\over T_{reh}}]_{(n=2)} $ & $8.13$ & $2.53$ & $1.062$ & $0.585$ \\
  $[{\sigma(T_{reh})\over T_{reh}}]_{(n=4)} $ & $13.64$ & $4.26$ & $1.044$ & $0.57$ \\
  \hline
\end{tabular}
\end{table}
Note that in the last column the relative uncertainty is less than
one. This could occur in the context of  WMAP9 results provided that
$\sigma(n_s)\leq 0.003$.

To compute the above uncertainties, a first order Taylor expansion
was employed (see (\ref{2000})) which, because of exponential
dependence of temperature on ${1\over 1-n_s}$, is insufficient. To
obtain more accurate bounds for $T_{reh}$ one can insert $n_s$ and
$\mathcal{P}_s$  directly in (\ref{48}). For example for quadratic
potential and using Planck+WP+highL+BAO data, at 2 sigma error
($\%95 CL$), we obtain
\begin{equation}
6.12\times 10^{11}GeV< T_{reh} <1.04\times 10^{15}GeV.
\end{equation}

\section{Conclusion}
Non-minimal derivative coupling model with a power law potential was
employed to describe the inflation (see (\ref{5})). In this context,
the slow roll inflationary phase of the inflaton was discussed in
high friction regime (see (\ref{11})). The rapid oscillation phase
after the slow roll, during which the inflaton decays to
ultra-relativistic particles, was studied. From the beginning of
this rapid oscillation until the radiation dominated epoch, the
equation of state parameter of the universe can be approximated by a
constant (see (\ref{r9})), which enabled us to compute the number of
e-folds in this era (see (\ref{28})). We also estimated the number
of e-folds number in radiation dominated, and recombination era, in
the same way as the minimal model (see (\ref{38})). By gathering all
these results together, we obtained the reheating temperature in
terms of $T_{CMB}$, spectral index and the amplitude of the power
spectrum of scalar perturbations (see (\ref{41})). Finally according
to recent astrophysical data, we determined the value of $T_{reh}$
which is much bigger than the reheating temperature obtained in
minimal model. This is due to high friction regime adopted in this
paper, which enhances the decay of the inflaton. We also showed that
the uncertainty in our result is very smaller with respect to the
minimal model.

Due to the transparency of the universe to gravitational wave  the
detection of primordial gravitational wave may also be used as a
powerful tool to study the history of the universe evolution.
Expansion rate of the universe and his thermal history after the
inflationary phase have direct imprints on the gravitational wave
spectrum and its detectability \cite{seto} . In \cite{grav},
depending on the tensor-to-scalar ratio $r$, and the reheating
temperature, the required sensitivity of some experiments to detect
gravitational wave was discussed. Indeed there was shown that $r$
and the reheating temperature, $T_{reh}$, are the main parameters
for determining the gravitational wave spectrum. As an outlook, in
future studies, these results may be extended to our non-minimal
derivative coupling model for which $r$ was computed in \cite{end}.


\begin{thebibliography}{99}

\bibitem{inf}A. H. Guth, Phys. Rev. D 23, 347(1981).
\bibitem{lind}A. Linde, Phys. Lett. B 129, 177 (1983); A.Linde, Particle Physics and Inflationary Cosmology (Harwood, Chur,
Switzerland, 1990).
\bibitem{Higgs} F. L. Bezrukov and M. E. Shaposhnikov, Phys. Lett. B 659, 703
(2008); R. N. Lerner and J. McDonald, Phys. Rev. D 82, 103525
(2010); R. N. Lerner and J. McDonald, Phys. Rev. D 83, 123522
(2011); K. Nakayama and F. Takahashi, JCAP 1102, 010 (2011),
arXiv:1008.4457 [hep-ph]; F. Bezrukov, A. Magnin, M. Shaposhnikov,
and S. Sibiryakov, JHEP 01, 016 (2011).
\bibitem{germ1}C. Germani and A. Kehagias, Phys. Rev. Lett. 105, 011302
(2010).
\bibitem{germ2}C. Germani and Y. Watanabe, JCAP 07(2011)031; C. Germani and A. Kehagias, arXiv:1003.4285 [astro-ph.CO].
\bibitem{germ3} C. Germani and A. Kehagias, Phys. Rev. Lett. 106, 161302
(2011);  C. Germani, arXiv:1112.1083v1 [astro-ph.CO].
\bibitem{non}S. V. Sushkov, Phys. Rev. D 80, 103505 (2009); E. N.
Saridakis and S. V. Sushkov, Phys. Rev. D 81, 083510 (2010); L. N. Granda, arXiv:1104.2253 [hep-th]; S. Tsujikawa,
arXiv:1201.5926v1 [astro-ph.CO]; H. M. Sadjadi, Phys. Rev. D 83,107301 (2011); G. Gubitosi and E. V. Linder, Phys. Lett. B 703, 113
(2011); A. Banijamali and B. Fazlpour, Phys. Lett. B 703, 366
(2011); S. A. Appleby, A. De Felice, and E. V. Linder, JCAP 1210 (2012) 060, arXiv:1208.4163 [astro-ph.CO];
S. V. Sushkov, Phys. Rev. D85, 123520 (2012).
\bibitem{reh}A. Albresht, P. J. Steinhardt, M. S. Turner,
and F. Wilczek, Phys. Rev. Lett. 48, 1437 (1982); L. Abbott, E.
Farhi, and M. Wise, Phys. Lett. B 117, 29 (1982); A. Doglov and
A.Linde, Phys. Lett. B 116, 329 (1982);
L. Kofman, A. D. Linde and A.A. Starobinsky, Phys. Rev. Lett 73, 3195 (1994);
L. Kofman, A. D. Linde and A.A. Starobinsky, Phys. Rev. D 56, 3258 (1997).
\bibitem{BBN}S. Hannestad, Phys. Rev. D 70, 043506 (2004).
\bibitem{Martin}J. Martin and C. Ringeval, Phys. Rev. D 82, 023511 (2010).
\bibitem{Mielc}J. Mielczarek, Phys. Rev. D 83, 023502 (2011).
\bibitem{WMAP}E. Komatsu et al., Astrophys. J. Suppl. 192, 18 (2011), arXiv:1001.4538 [astro-ph.CO].
\bibitem{large}A. R. Liddle and D. H. Lyth, Cosmological Inflation and Large-Scale Structure
(Cambridge University Press 2000); A. R. Liddle and D.H. Lyth,
Phys. Rept. 231, 1 (1993).
\bibitem{rapid} Y. Shtanov, J. Traschen, and R. Brandenberger, Phys. Rev. D 51, 5438
(1995); M. S. Turner, Phys. Rev. D 28, 1243 (1983).
\bibitem{god}H. M. Sadjadi and P. Goodarzi, JCAP 1302, 038 (2013), arXiv:1203.1580v2 [gr-qc].
\bibitem{kolb} E. Kolb and M. Turner, The Early Universe (Addison-Wesley Publishing
Company, Redwood City, California, 1990).
\bibitem{table} P. A. R. Ade et al., Planck 2013 results. XVI, arXiv:
1303.5076 [astro-ph] (2013); P. A. R. Ade et al., Planck 2013
results. I, arXiv: 1303.5062 [astro-ph) (2013); G. Hinshaw et al.
[WMAP Collaboration], arXiv:1212.5226 [astro-ph.CO]; C. L. Bennett
et al., arXiv:1212.5225[astro-ph.CO].
\bibitem{seto}N. Seto and J. Yokoyama, J. Phys. Soc. Jap. 72, 3082
(2003);
\bibitem{grav}K. Nakayama, S. Saito, Y. Suwa, and J. Yokoyama, JCAP 0806, 020
(2008).
\bibitem{end}C. Germani and A. Kehagias (NTUA, Athens), JCAP 1005, 019
(2010).
\end{thebibliography}
\end{document}